\documentclass{egpubl}
 \WsSubmission    % uncomment for submission to EG Workshop
\usepackage{graphicx} 
\PrintedOrElectronic

\usepackage{t1enc,dfadobe}

\let\it=\itshape

\title[Technical Note: Generating Realistic Fighting Scenes by Game Tree]%
{Technical Note: Generating Realistic Fighting Scenes by Game Tree}

\author[H. P. H. Shum, T. Komura]
{Hubert P. H. Shum and Taku Komura \\
	Institute of Perception, Action and Behavior,  School of Informations, University of Edinburgh \\
	Project Webpage  http://homepages.inf.ed.ac.uk/tkomura/fighting.html
}
\begin{document}
	\maketitle
	\thispagestyle{empty}
	
	\begin{abstract}
		Recently, there have been a lot of researches to synthesize / edit the motion of a single avatar
		in the virtual environment.  However, there has not been so much work of simulating continuous
		interactions of multiple avatars such as fighting.
		In this paper, we propose a new method to generate a realistic fighting scene based on
		motion capture data. We propose a new algorithm called the temporal expansion approach which
		maps the continuous time action plan to a discrete causality space such that turn-based
		evaluation methods can be used.
		As a result, it is possible to use many mature algorithms available in
		strategy games such as the Minimax algorithm and $\alpha-\beta$ pruning.
		We also propose a method to generate and use an offense/defense table, which illustrates the
		spatial-temporal relationship of attacks and dodges, to incorporate tactical maneuvers of defense
		into the scene.
		Using our method, avatars will plan their strategies taking into account the reaction of the
		opponent. Fighting scenes with multiple avatars are generated to demonstrate the effectiveness
		of our algorithm.  The proposed method can also be applied to
		other kinds of continuous activities that require strategy planning such as sport games.
		
		\begin{classification} % according to http://www.acm.org/class/1998/
			\CCScat{I.3.3}{Motion Control}{Motion Editing,}{Interactive Techniques for Specifying Motion,}{Inverse Kinematics}
		\end{classification}
	\end{abstract}
	
	\section{Introduction}
	Fighting is an event of continuous interactions of humans, that is difficult
	to be simulated on the computer.
	It involves various characteristics of humans, such as power, perception, and intelligence.
	When suddenly a person is attacked, he/she will try to get away from the
	offender by all means and will either try to run away or hit
	back according to the condition. After the enemy repeats the same attack
	a few times, a human will learn the pre-action of that attack and figure out
	an effective counter-attack.  Therefore, the fighting style and strategies will change as
	the match goes on. As a result, modeling the planning process of fighting requires
	various techniques of artificial intelligence and game theory.
	
	%Fighting can be categorized into  (1) matches of martial arts (2) fights/battle among
	%a group of people, and (3) a crowded scene in which there are pushes and pullings such as
	%in chaotic scenes such as fires. There is a great demand for generating and
	%simulating such kind of scenarios.
	There is a great demand for creating scenes of fighting in the movie, television and
	game industry.  For designing such a scene, there are people called
	"fighting choreographers", who plan how the fighting scene should be carried on.
	There are also professional extras that are
	specialized in fighting scenes so that they can be hit and blown in a realistic way.
	If the scene involves thousands of people, this can be even more complicated, as the
	scene must proceed in a certain context while at every location, each character must be
	seriously fighting with the enemy based on their own tactics.
	For games or animations, the animators usually manually design the motions themselves
	and combine several characters to generate the scene.
	In either of the above cases, the cost and time required for creating the scene
	can be enormous.  It will be far better if the computers can be used to simulate such
	scenes automatically.
	
	Generating an artistic scene is a difficult task as the criteria for beauty is difficult to
	explain. Instead, in this research, we will assume that a well-planned, serious fight between
	characters will make the scene realistic.
	In this paper, we propose a new algorithm called the temporal expansion approach which
	maps the continuous time action plan to a discrete causality space such that turn-based
	evaluation methods can be used.
	As a result, it is possible to use many mature algorithms available to do
	strategy planning such as the Minimax algorithms or $\alpha-\beta$ pruning.
	
	In addition to this, we propose to use an offense / defense table to let the avatars select
	the appropriate defense action to counteract the offense action by the opponent.
	In this table, for each entry of the attack, the appropriate
	defense motions together with the best timing to launch them are listed.
	The offense / defense table can be generated by capturing the motions of
	two people fighting with each other, and generating a histogram of the launched defense
	motions for each attack.
	
	Using our method, avatars will plan their strategies taking into account the reaction of the
	opponent. By editing the parameters of the criteria for fighting, it is possible to
	simulate various fighting styles, such as being more passive, aggressive, or preferring
	kicks than punches.  By increasing the depth of the game tree, it is possible to make the
	avatar to be more intelligent.  The methodology can be used not only for fighting but also for
	other continuous activities such as dancing or sport games.
	
	This document provides supplementary technical information for \cite{shum06generating}. We refer the readers to \cite{shum06generating} for further details.
	
	\section{Related Work}
	Motion editing and synthesis has become a huge research area which has many applications
	in computer graphics, robotics and biomechanics.  Most of the researches assume there
	is a set of motion data obtained using a motion capture device. Recently, a lot of
	techniques to edit / retarget \cite{Gleicher::SIGGRAPH98,Lee::SIGGRAPH99,Abe::SCA04}
	or synthesize a new sequence of character's motion using precaptured
	motion data \cite{MotionGenerationFromExamples,Lee::SIGGRAPH02,MotionGraph,Kovar::SIGGRAPH04,
		Mukai::SIGGRAPH2005} have been proposed.
	The Motion Graph approach~\cite{MotionGenerationFromExamples,Lee::SIGGRAPH02,MotionGraph}
	is a method that can interactively reproduce continuous motions of characters
	based on a graph that is automatically generated from captured motion data.
	Since the Motion Graph produces a lot of edges and nodes without any context,
	it becomes difficult to control the character as the user wishes. Recently, therefore,
	works to resolve such problems by introducing a hierarchical structure have been
	proposed~\cite{Lau::SCA05,Kwon::SCA05}.
	
	Most of these researches handle characters in the scene individually,
	and need extensions to handle dense interactions, such as pushing or pulling, among
	several characters.
	For such kind of effects, methods that combine simulations and motion capture data
	are known to be effective \cite{Arikan::SCA2005,Zordan::SIGGRAPH2005}.
	In case of fighting scenes, however, the offended characters must not just be pushed away, but
	need to defend and counterattack.
	%As a result, fighting involves continues dense interactions
	%of characters.
	Although there are so much research work for motion editing
	and synthesis, there has been little work for generating a fighting scene that
	involves more than two characters.
	In the movie Lord of the Rings, the motions of background characters fighting with
	each other are generated by very sparse Motion Graphs ~\cite{Griggs::IEEReview03}.
	However, since the motions used are limited and the AI engine used is not smart enough,
	it can be applied only for background characters and not the main characters.
	
	The only academic work that is known for fighting is the work by Lee et al.~\cite{Lee::SCA04},
	which is to generate a scene of two boxers fighting with each other.
	In their work, the shadow boxing of a single boxer is captured
	for several minutes, and a Motion Graph is generated automatically out of it.
	Each boxer uses optimal path search to decide the best action to approach and hit each other.
	However, since the animations were based on singularly captured motions, the interactions
	of the characters were limited and different from real boxers.
	There is no context for defense motions, and therefore, to find out which defense motion
	is effective for which offense motion a precise collision detection has to be done for every motion.
	
	{\it Endorphin}~\cite{endorphin} by Naturalmotion Inc., can generate scenes of
	dense interactions of the characters.
	The interactions between the characters are, however, instantaneous,
	and not continuous. In order to generate a fighting scene,
	it is necessary that the characters continuously
	offend / defend each other.
	
	\section{Contribution of this paper}
	The main contribution of this paper is that we propose a methodology called
	temporal expansion approach
	to enable avatars to plan for the fighting.  By using the temporal expansion approach, the
	continuous nature of the fighting is converted to a discrete strategy planning problem, in which
	AI techniques developed for games such as chess can be applied.
	We also have enabled to simulate various styles of fighting by changing the value of the
	parameters composing the objective function of the offense/defense actions.
	
	%The second contribution is proposing to use an offense/defense table that is generated
	%from motion of two subjects captured at the same time, to evaluate the interactions
	%of the fighters.  By using this table, the close interactions during the fighting can be
	%generated without precise collision detections.
	%We also propose a method to generate the table by capturing the motion of two characters
	%at the same time and also edit the motions based on the entries saved in this table.
	\section{Methodology}
	In this session, the algorithm to generate the fighting scene is explained.
	We will follow the Motion Graph approach which is often used in avatar control today.
	\subsection{Data Acquisition}
	We have captured the motions of boxers and kick boxers conducting sparring with each
	other. They wore the boxing gloves and the sparring was done seriously.
	%The Motion Graphs of both characters are generated using this data. In addition to the Motion Graph,
	%a table that describes which defense motion is effective to get away from which offense
	Based on the captured motion, the offense/defense table,
	that describes which defense motion is effective to get away
	from which offense motion was generated.
	For example, head slips and parries are effective for getting away from
	jabs and straights while sway-back action is effective to get away from upper-cuts.
	Generating such database manually is a tedious and time-consuming work.
	Doing this automatically by collision detection of arbitrary
	offense and defense motion is inaccurate as subtle change of the motion can
	completely change the amount of damage to the avatar. As a result, there is a risk
	that even though the defense motion is effective for avoiding some specific attacks,
	it is not evaluated correctly.  Actually defense involves a lot of factors,
	not only the collision but also visionary information, changing the stiffness of the muscles,
	and also timing. A well trained fighter can adjust the motion in a very short time to counteract
	different attacks.
	As we capture two characters fighting, the subjects will always try
	to defend the offense of the other character at the best timing using the most appropriate action.
	As a result, not only the suitable defense but also the best timing to launch the action will
	be known.
	
	\subsection{Action-level state machine}
	Here we define the term "motion" as rawly captured human motion,
	and the term "action" as a meaningful segment of the motion we captured.
	In the field of fighting, an action can be an attack (such as a "left straight", "jab" or a "right kick"),
	a defense (such as "parries", "blocking" or "ducking") , a transition (such as "stepping to the left",
	"stepping forward" or "back step") or a combination of these three.
	In a captured sequence of motions, there are multiple actions.
	%Figure \ref{fig:fig1} shows a motion with a punch action and a dodge action.
	%\begin{figure}[htb]
	%  \centering
	%  \includegraphics[width=1.0\linewidth]{fig1.png}
	%  \caption{\label{fig:fig1}Multiple action in a single motion}
	%\end{figure}
	
\begin{figure}[htb]
		\centering
		\includegraphics[width=1.0\linewidth]{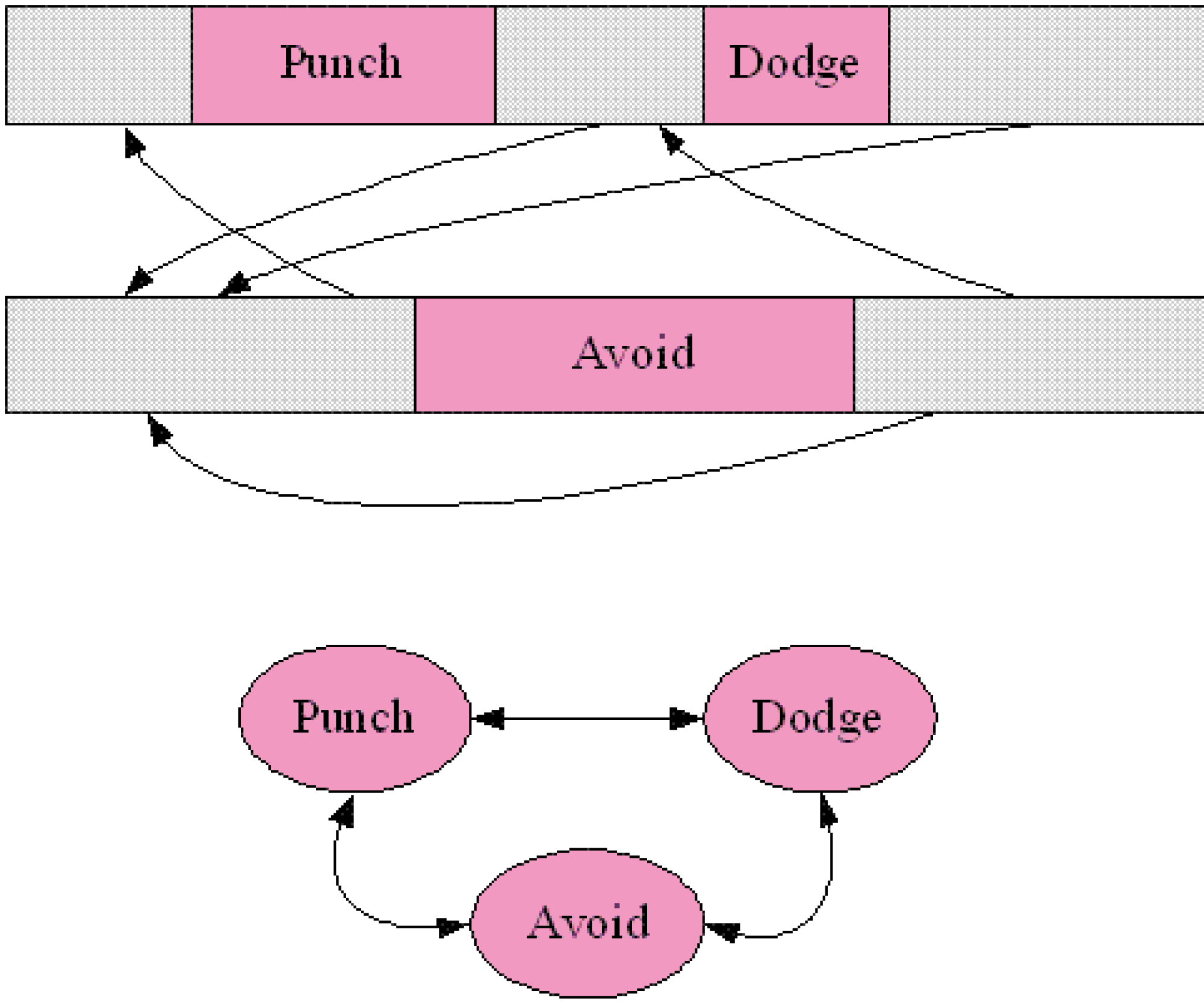}
		\caption{\label{fig:fig2}
			Frame level motion graph (upper) and action level state machine (lower)
			Same as the frame level motion graph, the action level motion graph is a strongly
			connected component \cite{MotionGraph} in a directed graph.}
	\end{figure}

	Using the captured motion, a basic frame-based Motion Graph~\cite{MotionGraph}
	is generated by automatically processing the data.
	However, such Motion Graph contains millions of nodes and edges and directly using such graph
	to plan the motion is inefficient.
	To speed up the planning process, an upper layer state machine is implemented upon the low level
	Motion Graph~\cite{Lau::SCA05,Kwon::SCA05}.
	%In order to emulate the thinking process of the fighters, the nodes in the upper layer must have a
	%logical meaning.
	We use the action defined above as the minimal unit in this upper layer.
	In the action level state machine, the smallest component is either an offense / defense motion or
	a transition motion to move from one location to another.  At this level, the underlying
	frame connectivity is hidden. Figure \ref{fig:fig2} illustrates the low level Motion Graph (upper)
	and the action level state machine (lower).
	Using the action-level state machine at the planning stage of the motion sequence will
	be similar to the human logical way of thinking, as people will always think of what
	sort of attack/defense/transition must be launched next during the match.
	Note that since the action is the smallest entity in the action level state machine,
	it cannot be subdivided into shorter actions. This means that once an
	action is started, it cannot be switched to another action in the middle,
	unless the opponent interrupts it by attacking.
	Once the Motion Graph of two layers are built, the shortest path to each actions are precomputed in the lower layer.
	Note that it is not necessary to compute the shortest path to all the nodes in the lower layer as in
	Lee et al.~\cite{Lee::SCA04},
	which will be very costly, but only to the nodes that correspond to the starting nodes of the
	actions.  As the number of actions are much limited than the number nodes in the lower layer,
	the computation for the shortest path is less costly.
	
	\subsection{Evaluating the actions}
	For selecting an action, there must be a measure to evaluate how much the avatar can
	benefit by launching each action. Such measure must take into account the current
	status of the avatar and opponent. With the evaluated results,
	the avatar can have a brief idea on whether such action is suitable or not.
	Such objective function can change according to the characteristics
	of the avatar or the style of fighting,
	However, basically they can be composed of three criteria:
	1. the location function that evaluates if the action is
	good in terms of the avatar's position and orientation,
	2. the fighter's characteristic function that evaluates how good the action is from
	the viewpoint of individual avatar, and
	3. the control function that evaluates the quality of the action from the viewer's point of view,
	such as the duration of the action and the frequency of usage.
	
	The location function is designed in a way that the avatar is in the
	reaching distance but not overlapping with the opponent.
	In addition to that, they should always face his/her opponent to prevent being attacked.
	
	The fighter's characteristic function evaluates how effective the action is in terms
	of fighting.
	In case of evaluating the attack, this function returns the score according to
	the effectiveness to knock down the opponent, and in case of defense,
	it returns how effective the motion is to reduce the effect of the attack by the opponent.
	
	The control function can be designed according to the application.
	If the designer prefers some specific good looking motions, they can favor those by
	giving higher marks to them when they are successful.
	We have tuned it to favor short and seldom used actions, as shorter motions will risk the
	body less, as the next motion can be launched shortly, and the use of seldom used actions
	ensures the variety of actions chosen.
	
	The outputs of the three criteria are combined together by calculating their weighted sum.
	As a result, the objective function can be written in the following form:
	\begin{eqnarray}
		\label{eq:objectivefunction}
		S_{\{A,B\}}(e_A, e_B, \Delta t, t_1, t_2) = w_l F^{loc}_{\{A,B\}} + w_f F^{fight}_{\{A,B\}}
		+ w_c F^{control}_{\{A,B\}}
	\end{eqnarray}
	where $S_{\{A,B\}}$ are the objective function for character A and B each,
	$e_A$ and $e_B$ are the action made by fighter A nd B, respectively, $\Delta t$ is the phase
	shift of the two actions, and $t_1\le t \le t_2$ is the range of time that the evaluation
	is done, $F^{loc}_{\{A,B\}}, F^{fight}_{\{A,B\}}, F^{control}_{\{A,B\}}$
	are the objective functions for location, fighting
	characteristics and control, respectively, and $w_l,w_f,w_c$ are their weight constants.
	The function for the two fighters can be different, and it is possible to change the
	characteristic of the fighter by tuning the values of the parameters.
	%Location functions are weighted by smaller values while field specific
	%functions are normally weighted by larger values. This is because the location functions are actually used to
	%support field specific functions. More specifically, the reason why an avatar moves towards the enemy
	%is to attack. The control functions are normally with small weight, since they are not related to the
	%content of the action and should not alter the resultant score too much.
	
	\subsection{Temporal Expansion}
	Using the objective function, we can evaluate how much an avatar can benefit by conducting
	different actions.
	However, for generating a realistic fighting scene, only considering the immediate
	benefit is not enough. Think of a computer-based algorithm to play chess.  A movement that
	shows the greatest
	effect in one ply (such as getting a valuable piece such as a castle or a bishop) is not necessarily
	the best choice at the end.  When deciding which movement is the best, AI algorithms
	expand the game tree and evaluate the static position after a few plies, to make a choice that
	benefits the player in long term.  Here we choose the same approach.
	\begin{figure}[htb]
		\centering
		\includegraphics[width=1.0\linewidth]{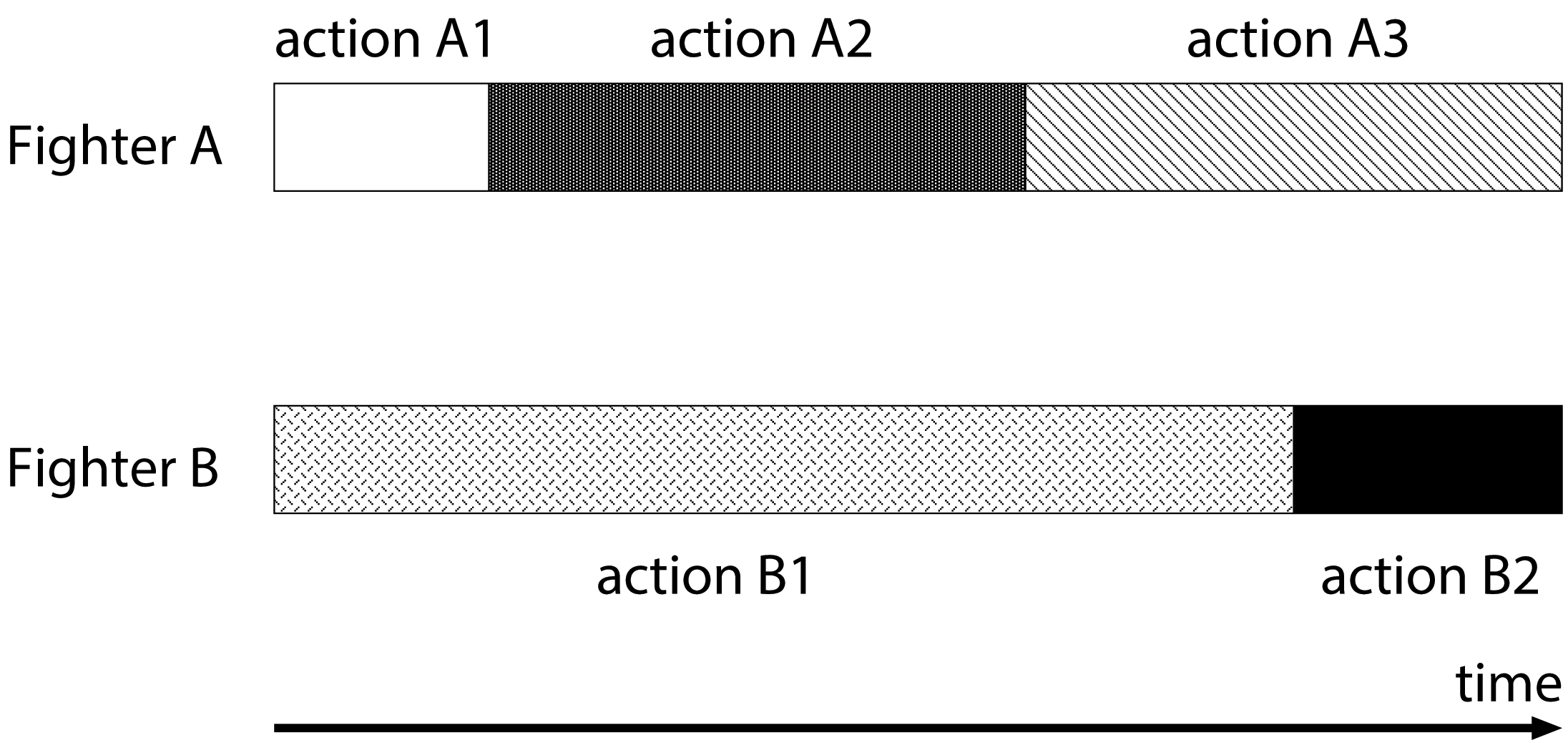}
		\caption{\label{fig:fighterAB}Motion sequence of two characters}
	\end{figure}
	The difference of chess and fighting is that the choice made by the players are not alternate,
	and this depends on the duration of the action done by each player.
	An example is illustrated in Figure \ref{fig:fighterAB}.
	Even though fighter A's second action $A2$ has started later than fighter B's current
	action $B1$, as $A2$ ends earlier than $B1$, fighter A needs to select and start the
	next action $A3$.
	%At the end of every action, the fighter chooses the next action to
	%conduct.
	%Once the action is decided, the shortest path from the current posture to the
	%next action in the MotionGraph will be looked up in the table and the corresponding action
	%will be conducted.
	%Every action is decided based on the temporal expansion of the game tree.
	
\begin{figure}[htb]
		\centering
		\includegraphics[width=1.0\linewidth]{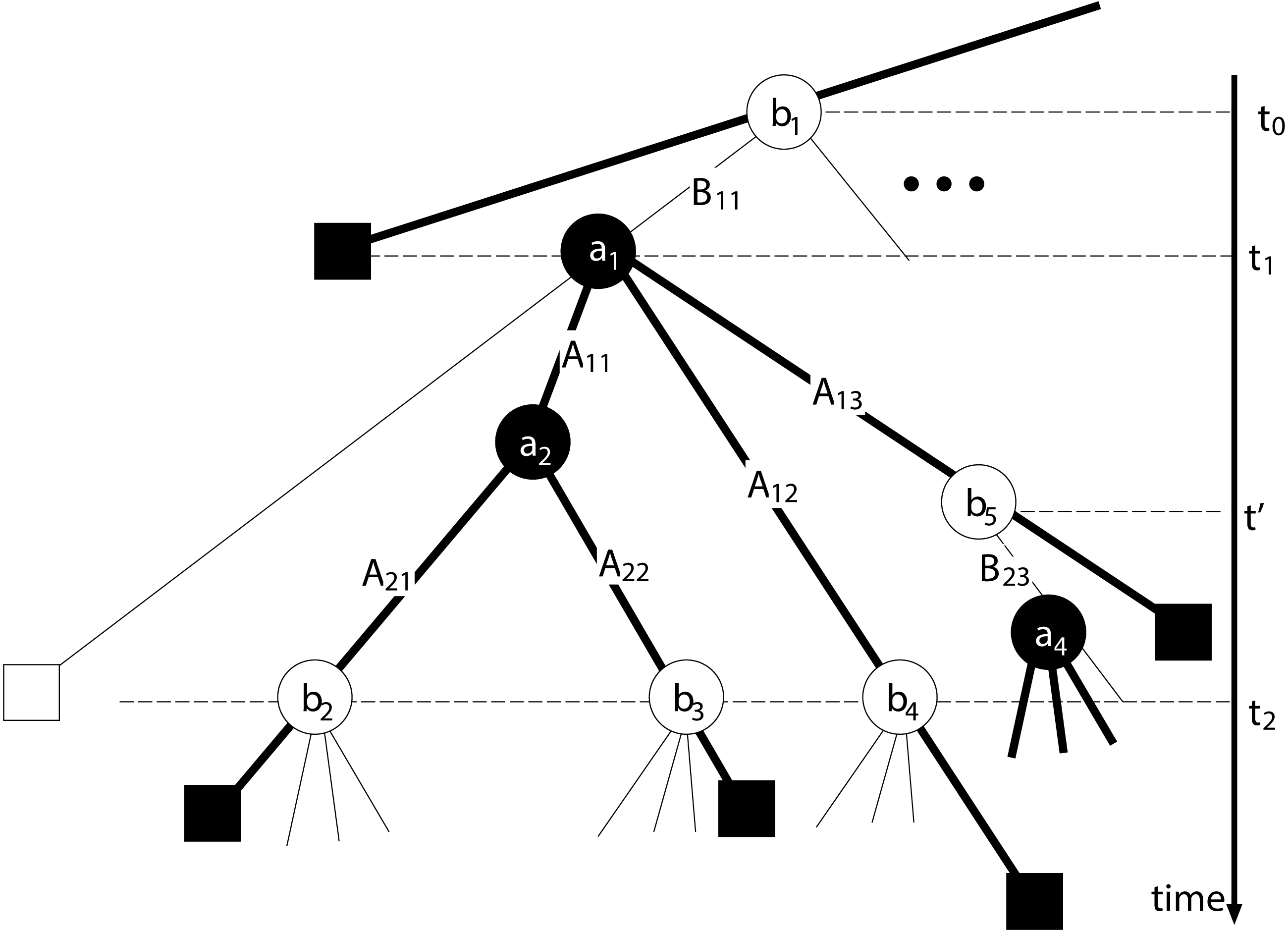}
		\caption{\label{fig:gametree}An expanded game tree of fighting. The distance along the vertical axis represents time.  The white and black circle nodes represent the moment fighter A and B launch new actions, respectively. Each edge represents the action that has been selected by the fighter and the square represents the end of each action}
	\end{figure}

	An example of the game tree that can handle this kind of non-alternate case
	is shown in Figure \ref{fig:gametree}.
	In this game tree, we assume that the depth along the vertical axis represents the time passed.
	The black and white circles represent the moment fighter A and B launching new actions, respectively.
	Each edge represents the action that has been selected by the fighter, and the square represents
	the moment that the action ends. The square is omitted and replaced with a circle node in case
	a successive motion is started at the same time by the same fighter.
	In this example the root node $b_1$ is the moment fighter B launches an action.
	The subtree shown is a game tree of fighter B selecting an action tagged $B_{11}$ at $t_0$.
	This action continues until time $t_2$.
	%as the white square on the left represents
	%the termination point.
	Since fighter A finishes the previous action at time $t_1$,
	which is before $B_{11}$ ends,
	a new node $a_1$ that represents the start of an action by fighter A is
	inserted into the tree. Say there are three actions $A_{11}$, $A_{12}$, and $A_{13}$ launchable
	at $a_1$. Since $A_{11}$ is short enough to end before $B_{11}$ ends, fighter A again can start
	another action at $a_2$.  On the other hand, $A_{12}$, which is another child action at $a_1$,
	is long enough so that it continues until fighter B finishes $B_{11}$ at $t_2$ and starts
	a new motion at $b_4$.
	So as well for the actions such as $A_{21}$ and $A_{22}$ launched at $a_2$.
	%Therefore, new white nodes $b_2$ and $b_3$, which represent the initiation of action by fighter B,
	%are inserted at time $t_2$.
	
	In some case, the fighter's action might be barred by the opponent by being punched or
	kicked.  In such case, the fighter might be either knocked down onto the ground immediately, or
	just lose balance and walk a few steps to recover the balance and resume the
	fight.  In either case, the response motion will be decided based on the
	current state of the body and the impulse added to the body.
	This is happening at node $b_5$ at time $t'$ in Figure \ref{fig:gametree}.
	The attack $A_{13}$ by fighter A hits fighter B,
	and as a result, the action $B_{11}$ by fighter B terminates at this moment and a reaction
	$B_{23}$ of stepping aside starts at the newly inserted node $b_5$.
	Since we assume the avatar hit does not have
	a choice for the action, only one edge is going out from node $b_5$.
	
	When expanding the game tree, it is necessary to specify the time in the future up to when
	we make the prediction.  Say this time is defined by $t_f$. Once we reach this time limit,
	we stop inserting new nodes under the current edge and evaluate the route to proceed to the
	Minimax algorithm.
	
	\subsection{Minimax Algorithm}
	We first need to calculate the
	score of the route from each fighter's point of view. This is calculated as follows:
	\begin{eqnarray}
		S_{\{A,B\}}^l = \sum_{i} S_{\{A,B\}}^{l,i}
	\end{eqnarray}
	where $S_{\{A,B\}}^{l,i}$ is the function defined in Equation \ref{eq:objectivefunction}
	that evaluates the status of fighter A and B when route $l$ passes through the $i$-th edge from
	the root.
	
	For using the Minimax algorithm, it is necessary to define a zero-sum static
	evaluation function for the leave nodes of the game tree.
	Say that positive score represents the benefits of fighter A and negative
	score represent the benefits of fighter B.
	Then, a zero-sum function can be defined by
	\begin{eqnarray}
		f(l) = S_A^l - S_B^l.
	\end{eqnarray}
	Now we need to evaluate the node $n$ of the game tree.
	For the leave nodes, they can be calculated by
	\begin{eqnarray}
		F(n) = \left \{ \begin{array}{ll}
			max_{e_i \in E_n}  ( f(l(e_i)) ) & \textrm{$n$ is fighter A's node}  \\
			min_{e_i \in E_n}  ( f(l(e_i)) ) & \textrm{$n$ is fighter B's node}  \\
		\end{array} \right.
	\end{eqnarray}
	where $E_n$ is the group of edges which are going out from node $n$, and $l(e_i)$ is the
	route from the root that includes $e_i$.
	The score of the internal nodes can be iteratively calculated
	up to the root by
	\begin{eqnarray}
		F(n) = \left \{ \begin{array}{ll}
			max_{n_i \in N_n}  ( F(n_i) ) & \textrm{$n$ is fighter A's node}  \\
			min_{n_i \in N_n}  ( F(n_i) ) & \textrm{$n$ is fighter B's node}  \\
		\end{array} \right.
	\end{eqnarray}
	where $N_n$ is the group of child nodes of node $n$.
	This computation is same as the Minimax algorithm used for alternate games such as chess.
	Therefore, $\alpha-\beta$ pruning can be applied to skip evaluating some of the nodes
	in the tree.
	\subsection{Pruning}
	Since our action-level state machine has nearly 100 nodes, without suitable pruning,
	the computation overhead cannot be afforded.
	Therefore, to increase the search speed, a sub-optimal search is used by introducing
	tree pruning. To ensure the pruning gives minimal effect to search quality, we only
	prune the nodes with poor immediate effect. For example, in fighting, simply being hit without
	defending rarely gives a positive influence in the future, and suddenly showing the back to
	the opponent does not benefit the fighter at all.  In addition to this, motions that will cause
	the characters to penetrate into each other too much will also be pruned.
	%The method greatly increases the performance of the temporal expansion process.
	\section{Experimental Results}
	We have first captured the motions of boxers and kick boxers sparring with each other for
	several minutes and generated the offense/defense table, Motion Graph, and action-level state
	machine.
	%The Motion Graph is composed of ??? nodes, ??? edges, while the action-level
	%motion graph contains ??? actions, which include ??? attacks, ??? defenses, and ??? transition
	%motions.
	
	Based on this data, we have generated various sequence of two avatars fighting by
	changing the parameters of the system.
	
\begin{figure*}[htb]
		\centering
		\includegraphics[width=0.23\linewidth]{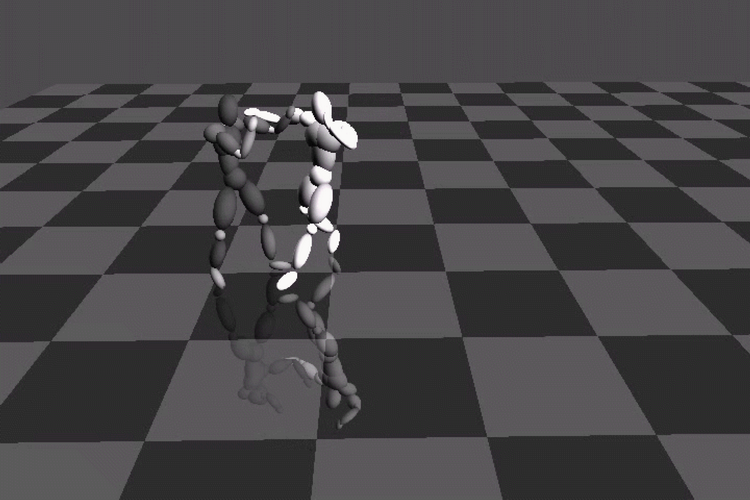}
		\includegraphics[width=0.23\linewidth]{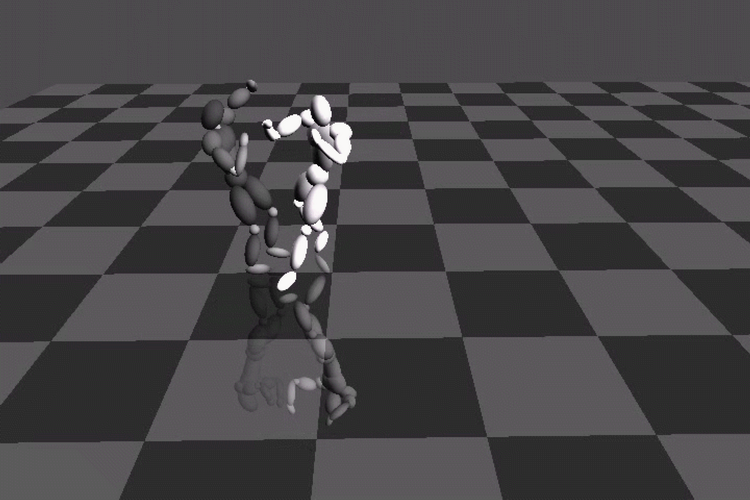}
		\includegraphics[width=0.23\linewidth]{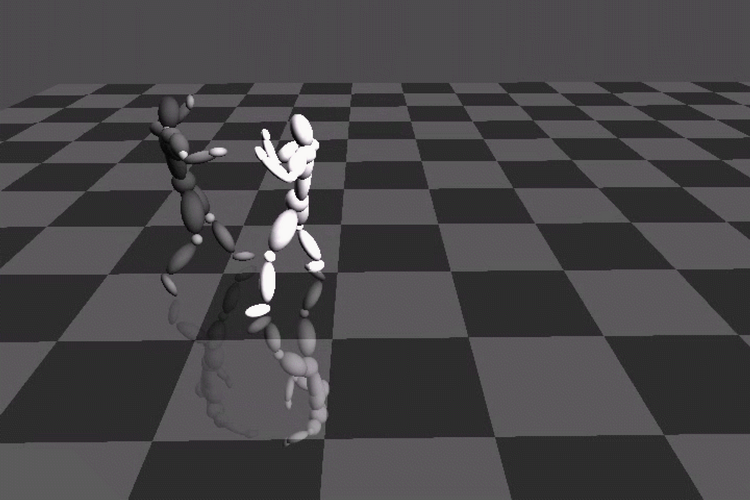}
		\includegraphics[width=0.23\linewidth]{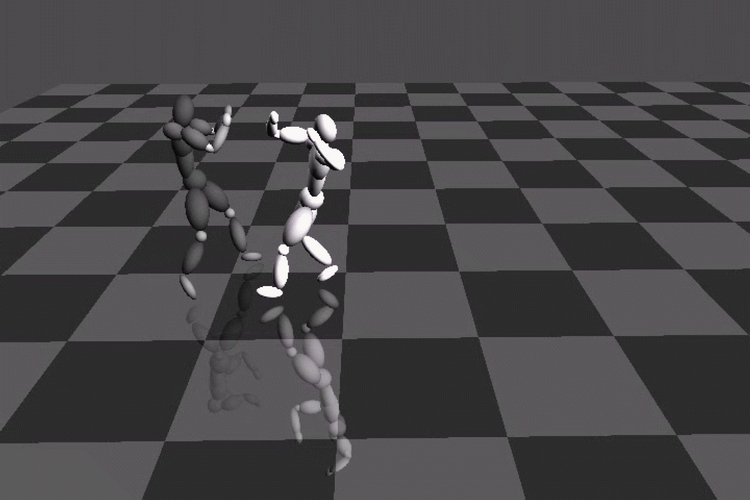}
		\caption{\label{fig:weakweak}
			A fight of avatars both with small $t_f$.  Since both avatars expand the tree only for a short time,
			they tend to chase short-term benefits and just hit each other.
		}
	\end{figure*}

	In the first experiment, we have limited the depth of the expansion only up to 1.0 seconds
	(Figure \ref{fig:weakweak}).
	Since most of the actions are longer than 1.0 second, both fighters will conduct actions to
	chase the immediate effect. As a result, both fighters tend to fight in a non-intelligent way.
	They both just launch attacks without caring about the risk of being counterattacked.
	As a result, an effect similar to young children quarreling can be obtained.
	
\begin{figure*}[htb]
		\centering
		\includegraphics[width=0.23\linewidth]{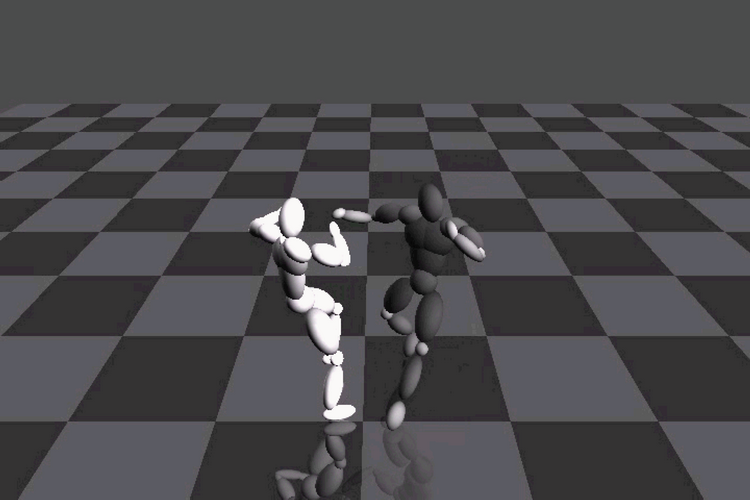}
		\includegraphics[width=0.23\linewidth]{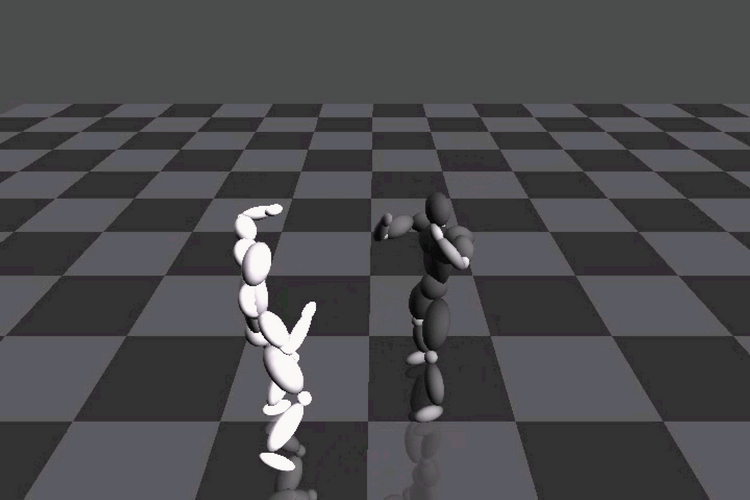}
		\includegraphics[width=0.23\linewidth]{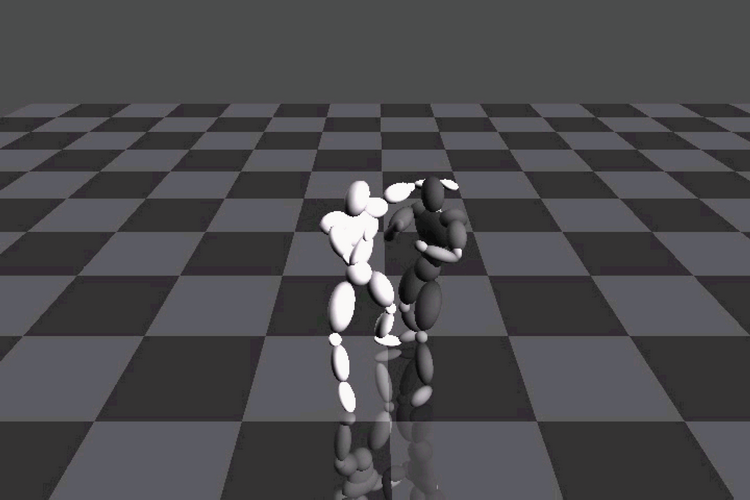}
		\includegraphics[width=0.23\linewidth]{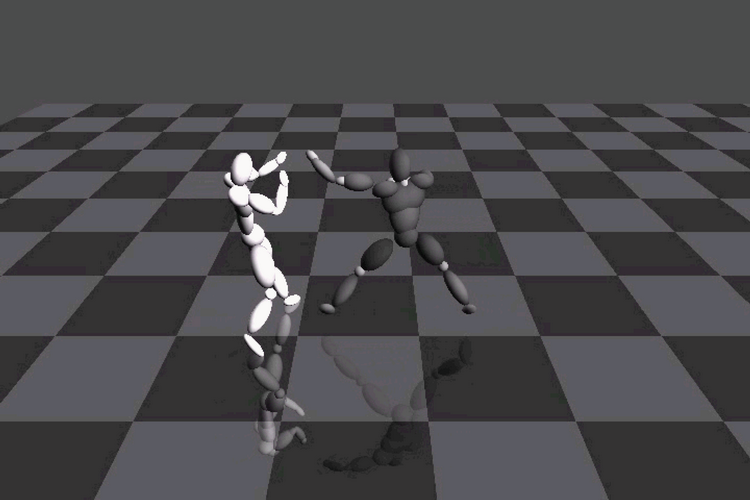}
		\caption{\label{fig:strongweak}
			A fight of avatars with small $t_f$ (white) and large $t_f$ (black).
			The avatar with small $t_f$ continues to get hit by the one with large $t_f$.
		}
	\end{figure*}

	In the second example, for one avatar, $t_f$ was increased to 3.0 seconds (Figure \ref{fig:strongweak}).
	Some of the snapshots are shown in \ref{fig:strongweak}.
	When simulating the fight, the avatar with larger $t_f$ will always succeed to hit the avatar with
	smaller $t_f$. Even though the avatar with smaller $t_f$ launches an attack, the opponent
	always succeed to defend and counterattack. Once a response motion to fall down is launched,
	the opponent will successively hit the avatar as it is a naive action.
	
\begin{figure*}[htb]
		\centering
		\includegraphics[width=0.23\linewidth]{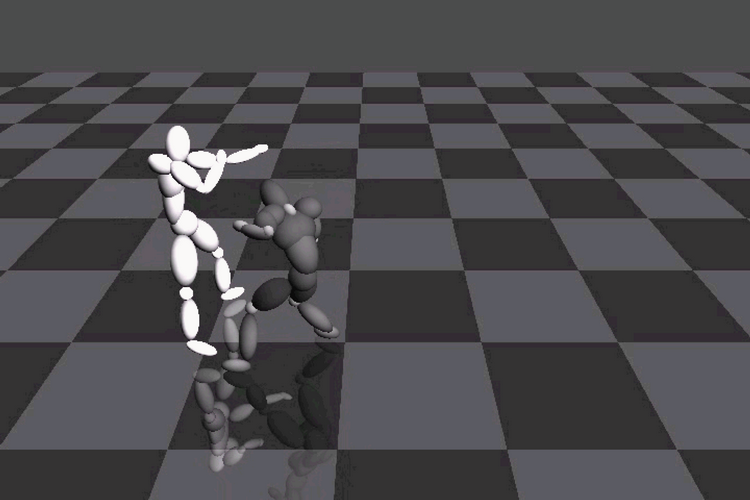}
		\includegraphics[width=0.23\linewidth]{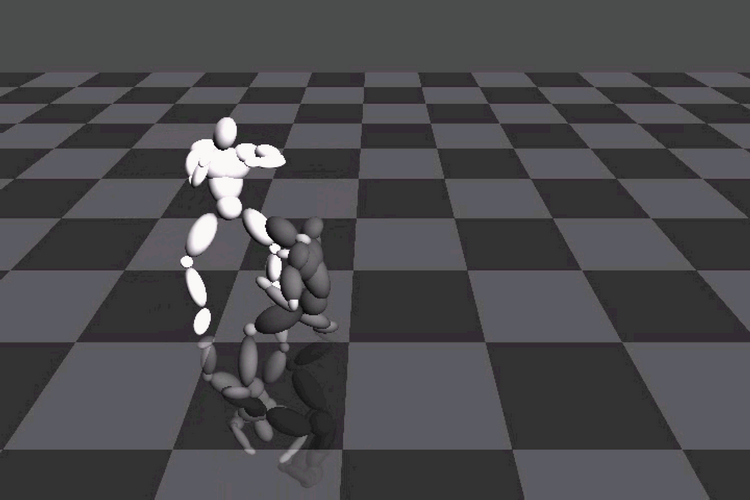}
		\includegraphics[width=0.23\linewidth]{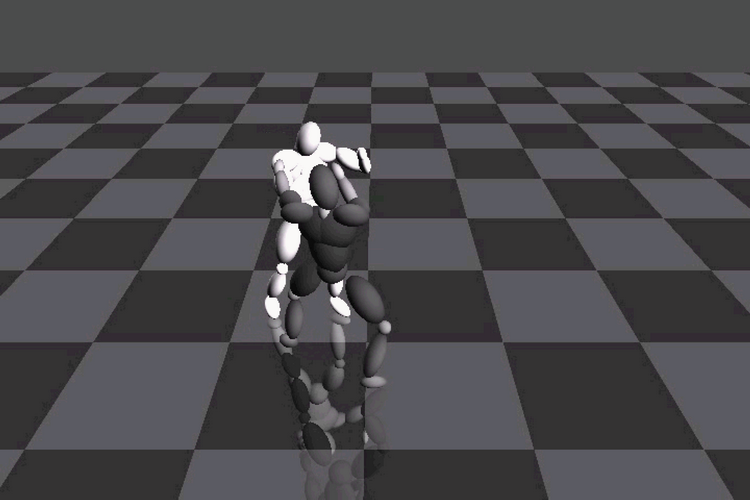}
		\includegraphics[width=0.23\linewidth]{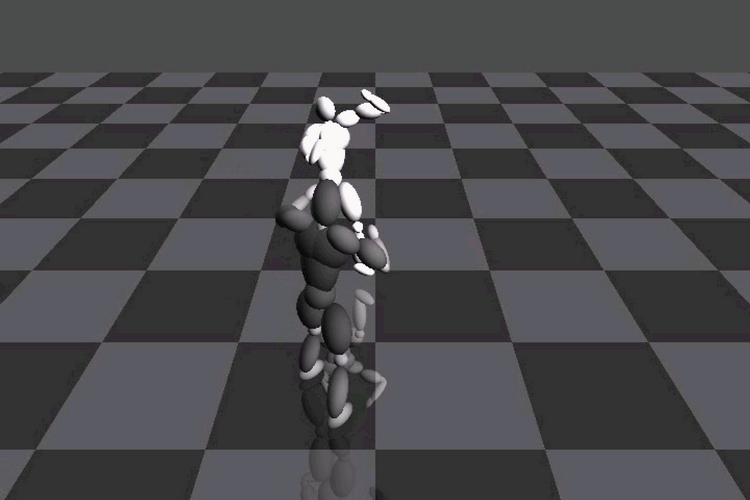}
		\caption{\label{fig:strongstrong}
			A fight of avatars both with large $t_f$.
			They tend to be more careful and defend more.
		}
	\end{figure*}

	Third, $t_f$ for both avatars were increased to 3.0 seconds (Figure \ref{fig:strongstrong}).
	In this example, both characters tend to defend much more than in the first example.
	Since defense motion will not give high marks to the defender in short term,
	they will not do it if $t_f$ is small.
	However, in long term, it certainly results in higher marks
	as they do not lose marks by being hit and the chance to counterattack increases.
	This time, both avatars become more careful, and motions to adjust their position also increases.
	However, the hits made by both are reduced, and can look less exciting for viewers who
	prefer to see more hits.
	
\begin{figure*}[htb]
		\centering
		\includegraphics[width=0.23\linewidth]{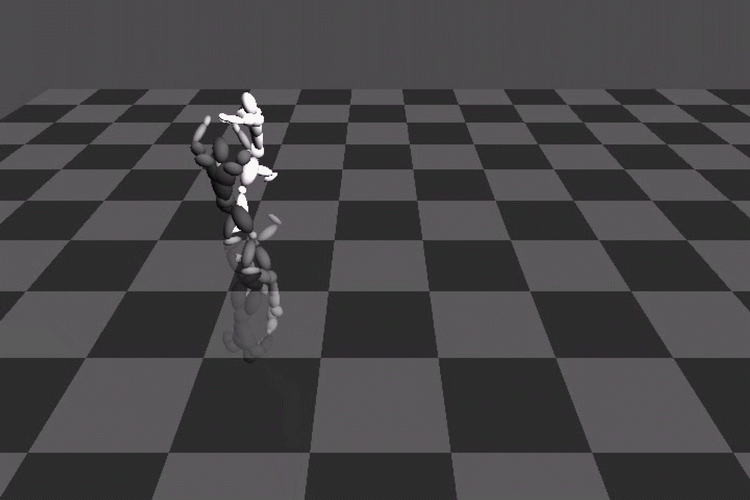}
		\includegraphics[width=0.23\linewidth]{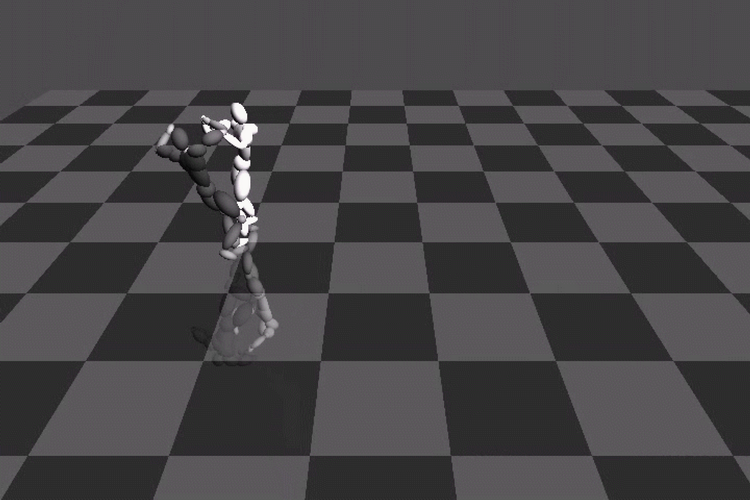}
		\includegraphics[width=0.23\linewidth]{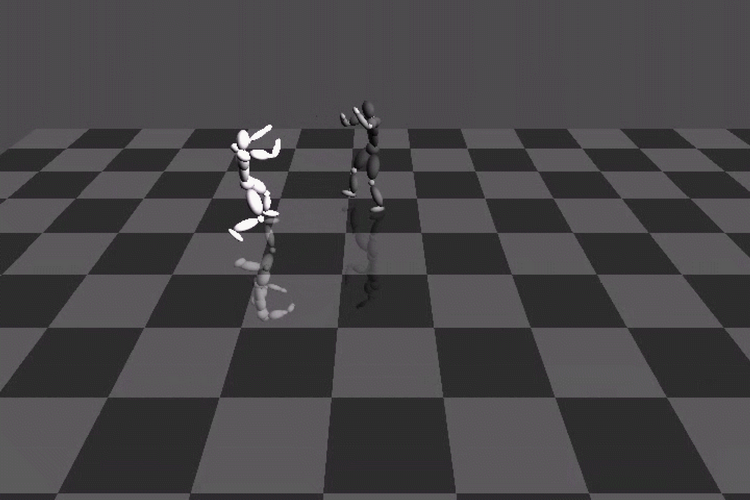}
		\includegraphics[width=0.23\linewidth]{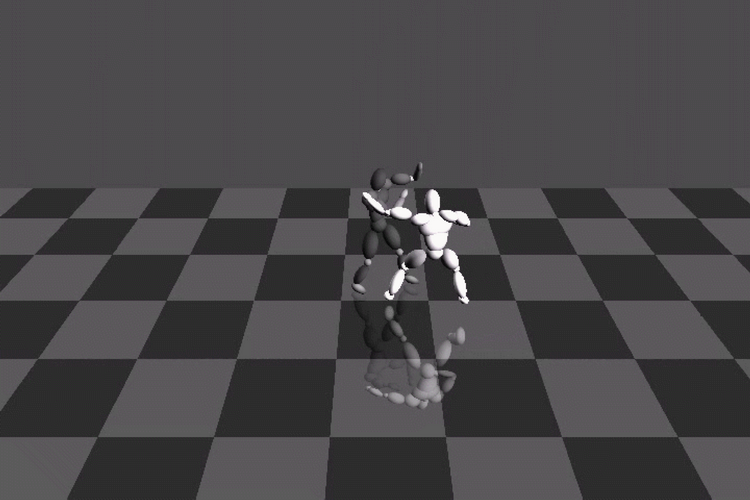}
		\caption{\label{fig:outboxer}
			A simulation of a fight between an outboxer (white) and an infighter (black).
			The outboxer is tuned so that he/she prefers longer distance and less interactions
			while the infighter is tuned in a way he/she prefers shorter distance and more interactions.
		}
	\end{figure*}

	In the final example, we have tuned the parameters of the fighter so that one shows the
	characteristics of an outboxer and the other of an infight boxer (Figure \ref{fig:outboxer}).
	The outboxer's location function was tuned so that he/she prefers longer distance and the
	characteristic function was tuned so that he/she prefers less interaction.
	On the other hand, the infighter's location function was tuned so that he/she prefers shorter
	distance and the characteristic function was tuned so that he/she prefers more interactions.
	In this example, the outboxer succeeds to always keep some distance, and sometimes suddenly
	steps in to hit the infight boxer.
	
	In some cases, the response of turning is a little bit slow comparing to people in the real world
	when having a match. This is due to the sparsity of the Motion Graph, and certainly other
	methodologies to fill in the sparsity of the motion
	space \cite{Mukai::SIGGRAPH2005,Kovar::SIGGRAPH04}
	must be introduced to generate a fully smooth fighting
	scene. However, the main contribution of our paper is on the temporal expansion approach.
	This methodology can be applied even though we combine this method with such interpolation
	techniques of the motion space.
	
	\section{Summary and Future Work}
	In this paper, we have proposed methods to generate realistic fighting scenes by using a new
	method called temporary expansion approach. Using our approach, it is possible to
	convert the problem of continuous time action planning to a
	discrete causality space problem such that turn-based evaluation methods can be used.
	We have shown various styles of fighting can be achieved by changing the parameters of the
	game tree such as its depth and the static evaluation function.
	The proposed method can be used for various applications such as computer games and animation of
	fighting scenes.
	
	In real fights, human always adjust such parameters during the fight according to their
	knowledge of the opponent, fatigue, and injuries. In the examples made in this paper, all such
	parameters were kept constant and it is assumed the opponent has full knowledge of the avatar.
	By dynamically adjusting such parameters according to the flow of the fight will increase the
	reality of the fight.
	
	Currently, the generation of the offense/defense table and the generation of the action-level
	state machine requires manual work, it is necessary to segment the action and categorize them
	into different groups. This requires a lot of time and knowledge of the style of fight. By
	using automatic segmentation techniques\cite{Jenkins::ICML04},
	it might be possible to further make this process automatic.
	
	\bibliographystyle{eg-alpha}
	\bibliography{biomech}

\newcommand{\etalchar}[1]{$^{#1}$}
\begin{thebibliography}{\uppercase{ZMCF05}}

\bibitem[AF02]{MotionGenerationFromExamples}
\textsc{Arikan O., Forsyth D.}:
\newblock Motion generation from examples.
\newblock \emph{ACM Transactions on Graphics 21}, 3 (2002), 483--490.

\bibitem[ALP04]{Abe::SCA04}
\textsc{Abe Y., Liu C.~K., Popovic Z.}:
\newblock Momentum-based parameterization of dynamic character motion.
\newblock \emph{Proceedings of ACM SIGGRAPH/Eurographics Symposium on Computer
  Animation} (2004), 173--182.

\bibitem[FO05]{Arikan::SCA2005}
\textsc{Forsyth O. A. D.~A., O'Brien J.~F.}:
\newblock Pushing people around.
\newblock \emph{Proceedings of 2005 ACM SIGGRAPH/Eurographics Symposium on
  Computer Animation} (2005), 59--66.

\bibitem[Gle98]{Gleicher::SIGGRAPH98}
\textsc{Gleicher M.}:
\newblock Retargetting motion to new characters.
\newblock \emph{Computer Graphicsi Proceedings, Annual Conference Series}
  (1998), 33--42.

\bibitem[Gri03]{Griggs::IEEReview03}
\textsc{Griggs K.}:
\newblock Assault on the senses [pc-run computer program for movies].
\newblock \emph{IEE Review 49}, 3 (2003), 24--27.

\bibitem[JM04]{Jenkins::ICML04}
\textsc{Jenkins O.~C., Matari\&\#263; M.~J.}:
\newblock A spatio-temporal extension to isomap nonlinear dimension reduction.
\newblock In \emph{ICML '04: Proceedings of the twenty-first international
  conference on Machine learning} (New York, NY, USA, 2004), ACM Press, p.~56.

\bibitem[KG04]{Kovar::SIGGRAPH04}
\textsc{Kovar L., Gleicher M.}:
\newblock Automated extraction and parameterization of motions in large data
  sets.
\newblock \emph{ACM Transactions on Graphics 23}, 3 (2004), 559--568.

\bibitem[KGP02]{MotionGraph}
\textsc{Kovar L., Gleicher M., Pighin F.}:
\newblock Motion graphs.
\newblock \emph{ACM Transactions on Graphics 21}, 3 (2002), 473--482.

\bibitem[KS05]{Kwon::SCA05}
\textsc{Kwon T., Shin S.~Y.}:
\newblock Motion modeling for on-line locomotion synthesis.
\newblock In \emph{SCA '05: Proceedings of the 2005 ACM SIGGRAPH/Eurographics
  symposium on Computer animation} (New York, NY, USA, 2005), ACM Press,
  pp.~29--38.

\bibitem[LCR{\etalchar{*}}02]{Lee::SIGGRAPH02}
\textsc{Lee J., Chai J., Reitsma P. S.~A., Hodgins J.~K., Pollard N.~S.}:
\newblock Interactive control of avatars animated with human motion data.
\newblock \emph{ACM Transactions on Graphics 21}, 3 (2002), 491--500.

\bibitem[LK05]{Lau::SCA05}
\textsc{Lau M., Kuffner J.~J.}:
\newblock Behavior planning for character animation.
\newblock In \emph{SCA '05: Proceedings of the 2005 ACM SIGGRAPH/Eurographics
  symposium on Computer animation} (New York, NY, USA, 2005), ACM Press,
  pp.~271--280.

\bibitem[LL04]{Lee::SCA04}
\textsc{Lee J., Lee K.~H.}:
\newblock Precomputing avatar behavior from human motion data.
\newblock \emph{Proceedings of 2004 ACM SIGGRAPH/Eurographics Symposium on
  Computer Animation} (2004), 79--87.

\bibitem[LS99]{Lee::SIGGRAPH99}
\textsc{Lee J., Shin S.~Y.}:
\newblock A hierarchical approach to interactive motion editing for human-like
  figures.
\newblock \emph{Proceedings of SIGGRAPH'99} (1999), 39--48.

\bibitem[MK05]{Mukai::SIGGRAPH2005}
\textsc{Mukai T., Kuriyama S.}:
\newblock Geostatistical motion interpolation.
\newblock \emph{ACM Trans. Graph. 24}, 3 (2005), 1062--1070.

\bibitem[Nat]{endorphin}
\textsc{Naturalmotion.Inc.}:
\newblock Endorphin.
\newblock \emph{http://www.naturalmotion.com}.

\bibitem[SK06]{shum06generating}
\textsc{Shum H. P.~H., Komura T.}:
\newblock Generating realistic fighting scenes by game tree.
\newblock In \emph{Proceedings of the 2006 ACM SIGGRAPH/Eurographics Symposium
  on Computer animation} (Aire-la-Ville, Switzerland, Switzerland, Sep 2006),
  SCA '06, Eurographics Association.

\bibitem[ZMCF05]{Zordan::SIGGRAPH2005}
\textsc{Zordan V.~B., Majkowska A., Chiu B., Fast M.}:
\newblock Dynamic response for motide capture animation.
\newblock \emph{ACM Transactions on Graphics 24}, 3 (2005), 697--701.

\end{thebibliography}
	
\end{document}